\documentclass{PoS}
\usepackage{graphicx}
\usepackage{dcolumn}
\usepackage{float}
\usepackage{amsmath}
\usepackage{amsfonts}
\usepackage{multirow}
\usepackage{color}
\usepackage{wasysym}
\newcommand{\lb}{\left(}
\newcommand{\rb}{\right)}
\def\GeV{\ifmmode {\mathrm{\ Ge\kern -0.1em V}}\else
                   \textrm{Ge\kern -0.1em V}\fi}%
\def\TeV{\ifmmode {\mathrm{\ Te\kern -0.1em V}}\else
                   \textrm{Te\kern -0.1em V}\fi}%
\def\bentarrow{\:\raisebox{1.3ex}{\rlap{$\vert$}}\!\rightarrow}
\def\bothdk#1#2#3#4#5{
\begin{array}{r c l}
#1 & \rightarrow & #2#3 \\
 & & \:\raisebox{1.3ex}{\rlap{$\vert$}}\raisebox{-0.5ex}{$\vert$}
\phantom{#2}\!\bentarrow #4 \\
 & & \bentarrow #5
\end{array}
}

\title{WW + jet at 14 and 100 TeV}

\ShortTitle{WW + jet at 14 and 100 TeV}
\author{John Campbell\\
        Fermilab\\ 
        E-mail: \email{johnmc@fnal.gov}}

\author{David Miller\\
        University of Glasgow\\ 
        E-mail: \email{david.j.miller@glasgow.ac.uk}}

\author{\speaker{Tania Robens}%
       \\
       IKTP, TU Dresden\\
       E-mail: \email{tania.robens@tu-dresden.de}}

\abstract{In the current LHC run, an accurate understanding of Standard Model processes is extremely important. Processes including electroweak gauge bosons serve as standard candles for SM measurements, and equally constitute important backgrounds for Beyond-the-Standard Model (BSM) searches. We present here the next-to-leading order (NLO) QCD virtual contributions to $W^+W^- +$jet in an analytic format obtained through unitarity methods. We present results for the full process using the Monte Carlo event generator MCFM, and discuss total as well as differential cross-sections for the LHC with 14 TeV center-of-mass energy, as well as a future 100 TeV proton-proton machine.

\hfill FERMILAB-CONF-16-514-T}

\FullConference{38th International Conference on High Energy Physics\\
                  3-10 August 2016\\
                 Chicago, USA}

\begin{document}
\section{Overview}
\begin{figure}
\begin{center}
\includegraphics[scale=0.3]{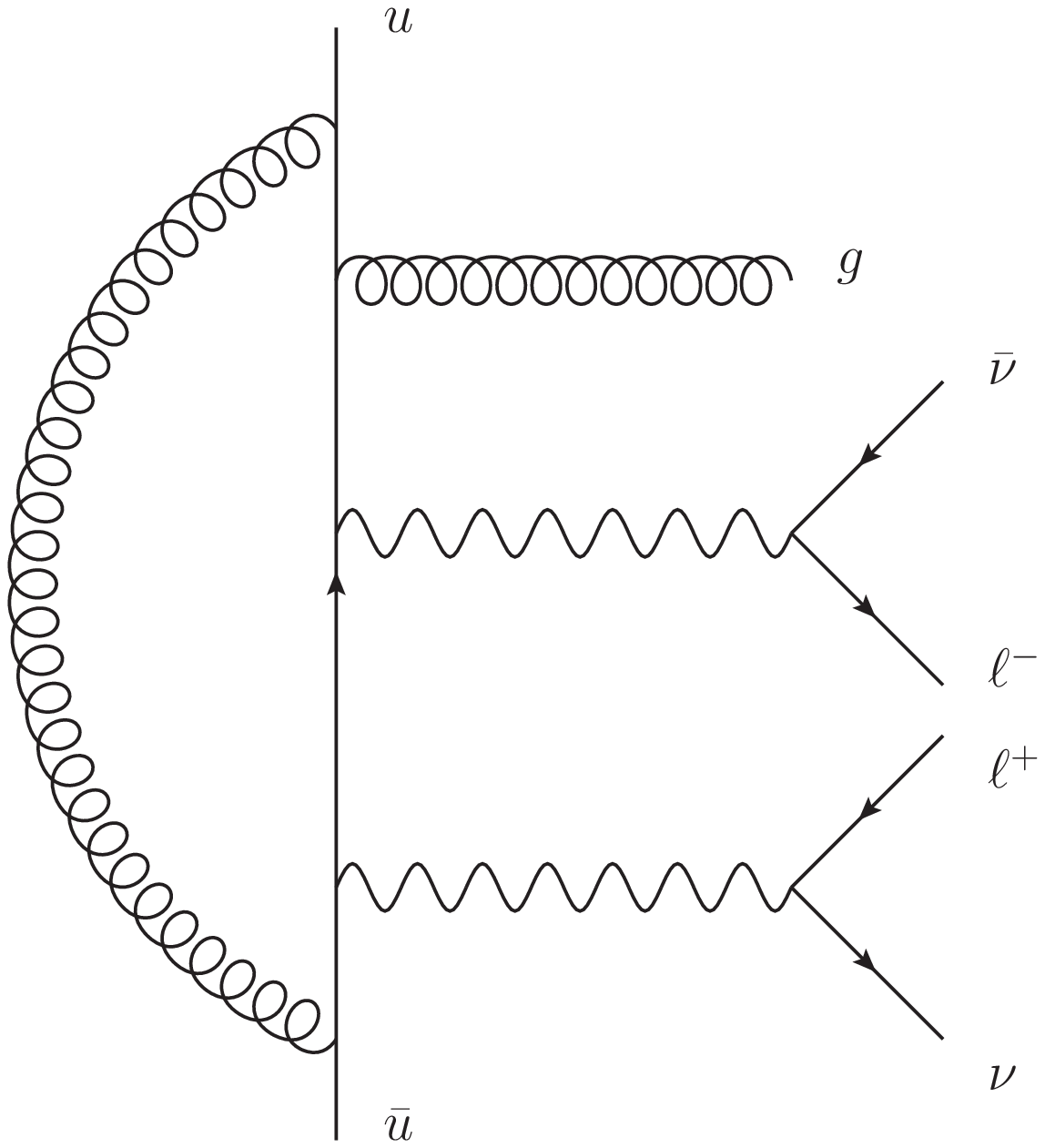} \hspace*{1.5cm}
\includegraphics[scale=0.3]{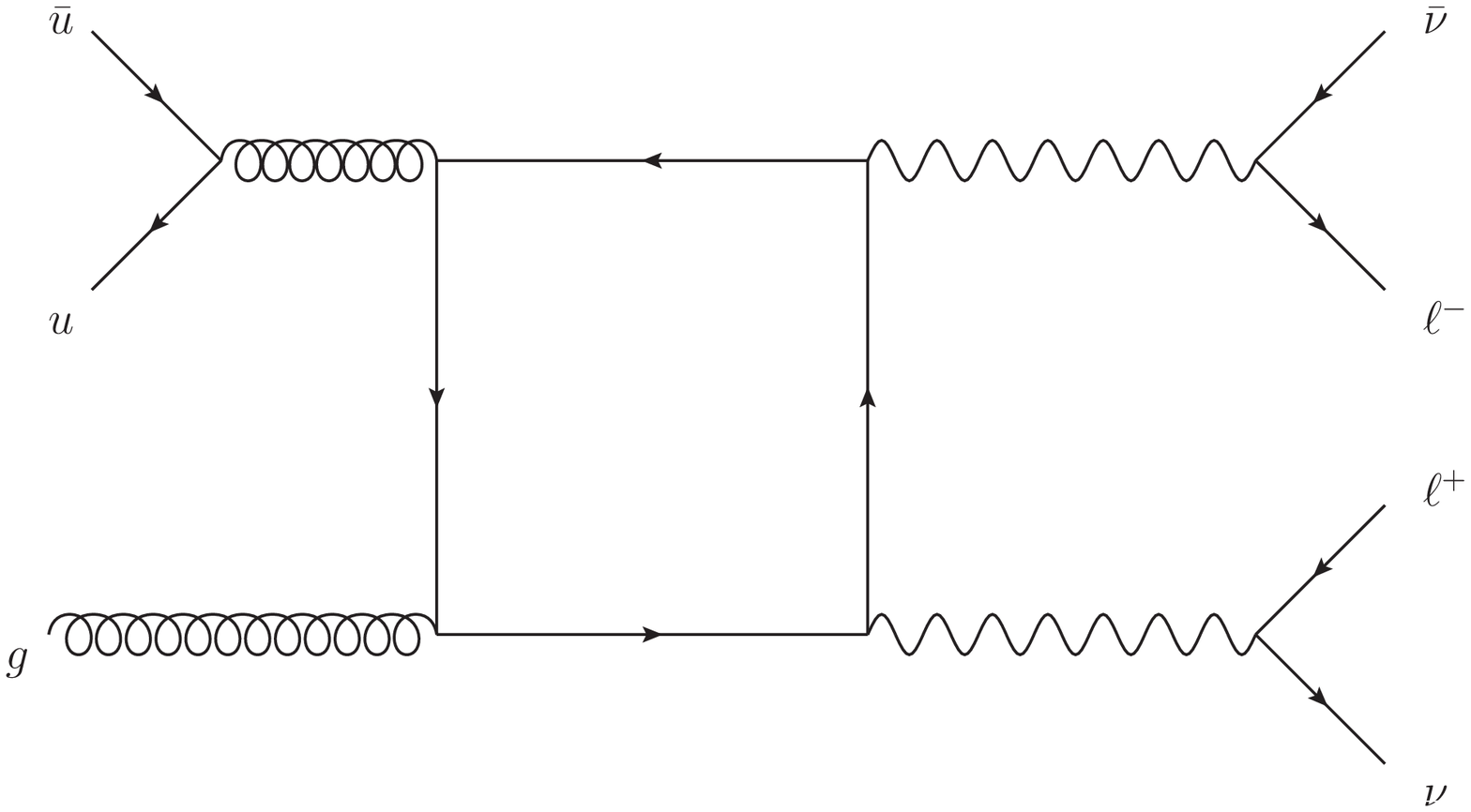}
\end{center}
\caption{Sample diagrams entering the calculation of the one-loop amplitude
for the $WW+$jet process.  The one-loop diagrams can be categorized according
to whether a gluon dresses a leading-order amplitude (left), or whether the
diagram includes a closed fermion loop (right).\label{fig:loopdiags}}
\end{figure}

We consider the hadronic production of $W$ pairs in association with a single
jet at next-to-leading order (NLO) in QCD at a hadron collider with a center-of-mass energy of 14 and 100 TeV. The $W$ bosons decay leptonically, with all spin correlations included.  At tree level
this process corresponds to the partonic reaction,
\begin{equation}
\label{WWjetprocess}
\bothdk{q+\bar q}{W^+ +}{W^-+g}{\mu^-+\nu_\mu}{\nu_e+e^+}
\end{equation}
with all possible crossings of the partons between initial and final states. Here, the $W$ bosons can either be radiated off a quark line or mediated via an offshell Z-boson that decays into a $W^+\,W^-$ pair. 
Next-to-leading order contributions include the emission of an additional parton, either as a virtual particle
to form a loop amplitude, or as a real external particle (cf. Fig.~\ref{fig:loopdiags}). All results presented here have been obtained using
Ref.~\cite{Campbell:2015hya}, where we made use of the methods of generalized unitarity \cite{Britto:2004nc,Britto:2005ha,Britto:2006sj,Forde:2007mi,Mastrolia:2009dr,Badger:2008cm}, furthermore employing the S@M Mathematica package~\cite{Maitre:2007jq} for the analytic treatment and simplification.
The evaluation of the scalar integrals has been performed using
the QCDLoop Fortran library~\cite{Ellis:2007qk}. The combination of the virtual contributions with born and real emission diagrams has been implemented
using MCFM~\cite{Campbell:1999ah,Campbell:2015qma}, and applied in a recent analysis by the ATLAS collaboration \cite{Aaboud:2016mrt}. Note that we do not include the effects of
any third-generation quarks.
\section{Analytic and numerical results}
Explicit expressions for sample coefficients have been presented in detail in \cite{Campbell:2015hya,Campbell:2016fph} and will not be repeated here. Instead, we choose to discuss results specific for a proton-proton collider with a center-of-mass energy of 100 TeV (see also \cite{Mangano:2016jyj}). Electroweak parameters as given in Tab.~\ref{parameters} were used for all results presented here.
\begin{table}
\begin{center}
\begin{tabular}{|c|c|c|c|}
\hline
$m_W$               & 80.385 GeV           & $\Gamma_W$ & 2.085 GeV \\
$m_Z$               & 91.1876 GeV          & $\Gamma_Z$ & 2.4952 GeV \\
$e^2$               & 0.095032             & $g_W^2$    & 0.42635 \\ 
$\sin^2\theta_W$    & $0.22290$            & $G_F$      & $0.116638\times10^{-4}$ \\
\hline
\end{tabular}
\caption{The values of the mass, width and electroweak parameters used. 
\label{parameters}}
\end{center}
\end{table}
In calculations of LO quantities we employ the CTEQ6L1 PDF set~\cite{Pumplin:2002vw},
while at NLO we use CT10~\cite{Lai:2010vv}.  The renormalization and factorization scales are
usually chosen to be the same, $\mu_R = \mu_F = \mu$, with our default scale choice
$\mu_0 \equiv \frac{H_T}{2} = \frac{1}{2} \sum_i p_{\perp}^i $.
The sum over the index $i$ runs over all final state leptons and partons.
Jets are defined using the anti-$k_T$ algorithm with separation parameter
$R=0.5$ and must satisfy
$p_{\perp}^\text{jet} > p_{\perp, \text{cut}}^{\text{jet}} \;,
|\eta^\text{jet}| <4.5 $.

\begin{figure}
\begin{center}
\includegraphics[scale=0.35,angle=-90]{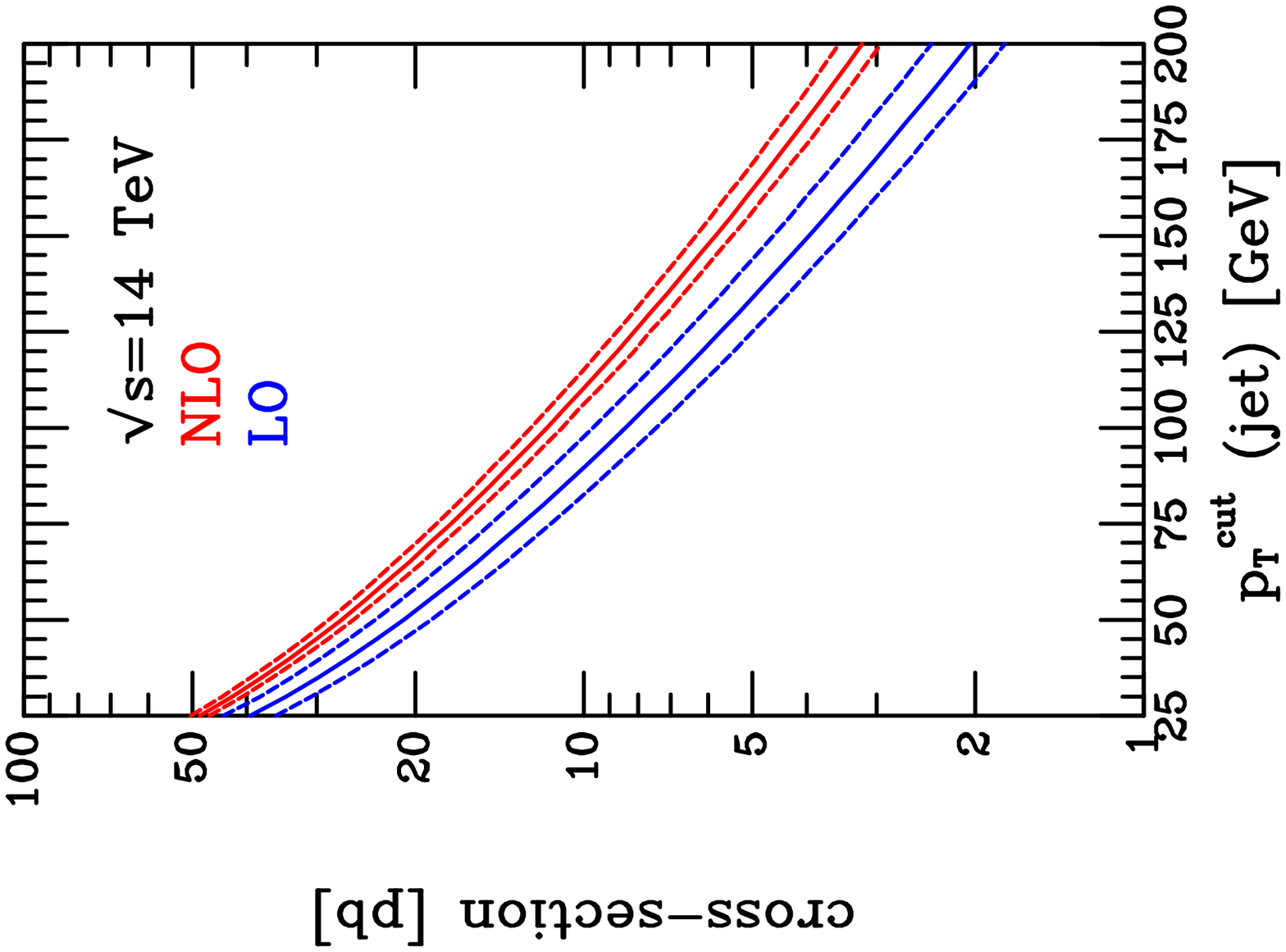} \hspace*{0.5cm}
\includegraphics[scale=0.35,angle=-90]{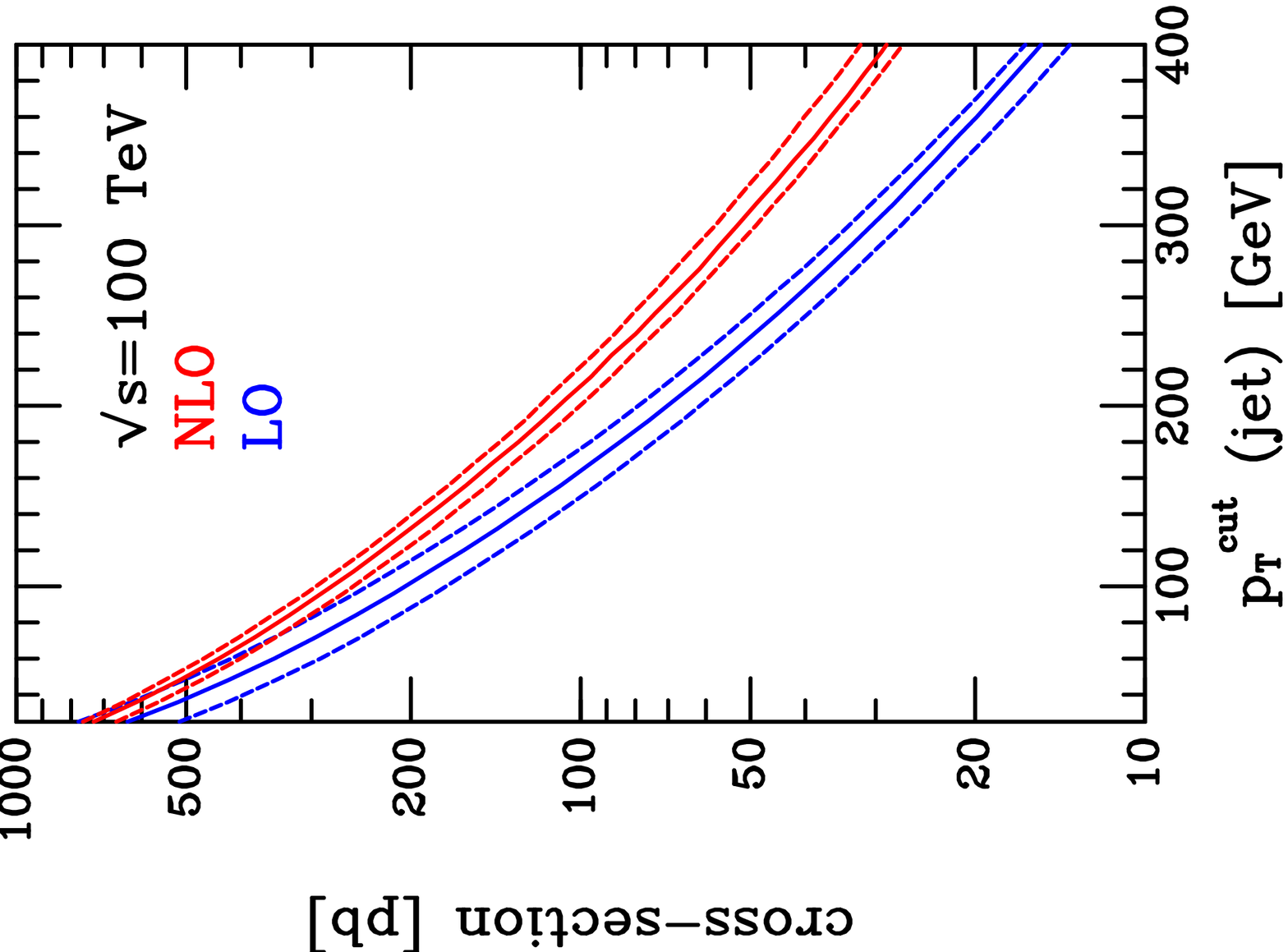}
\end{center}
\caption{Cross-sections at $\sqrt s = 14$~TeV (left) and $100$~TeV (right),
as a function of the transverse momentum
cut on the jet.  The prediction at each order is shown as a solid line,
with the dotted lines indicating the scale uncertainty corresponding to a factor of two variation
about the central scale. 
\label{fig:ptjet}}
\end{figure}

\begin{figure}
\begin{center}
\includegraphics[width=0.3\textwidth, angle=-90]{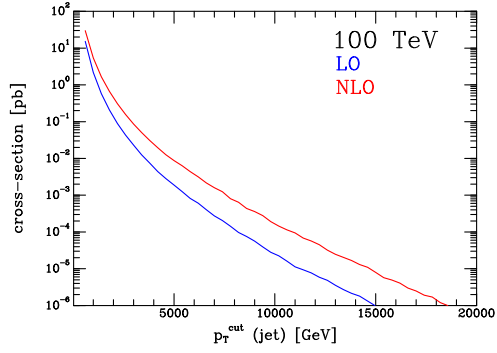}
\end{center}
\caption{Cross-sections at  $100$~TeV,
as a function of the transverse momentum
cut on the jet.
\label{fig:ptjet2}}
\end{figure}

\begin{figure}
\begin{center}
\includegraphics[width=0.3\textwidth, angle=-90]{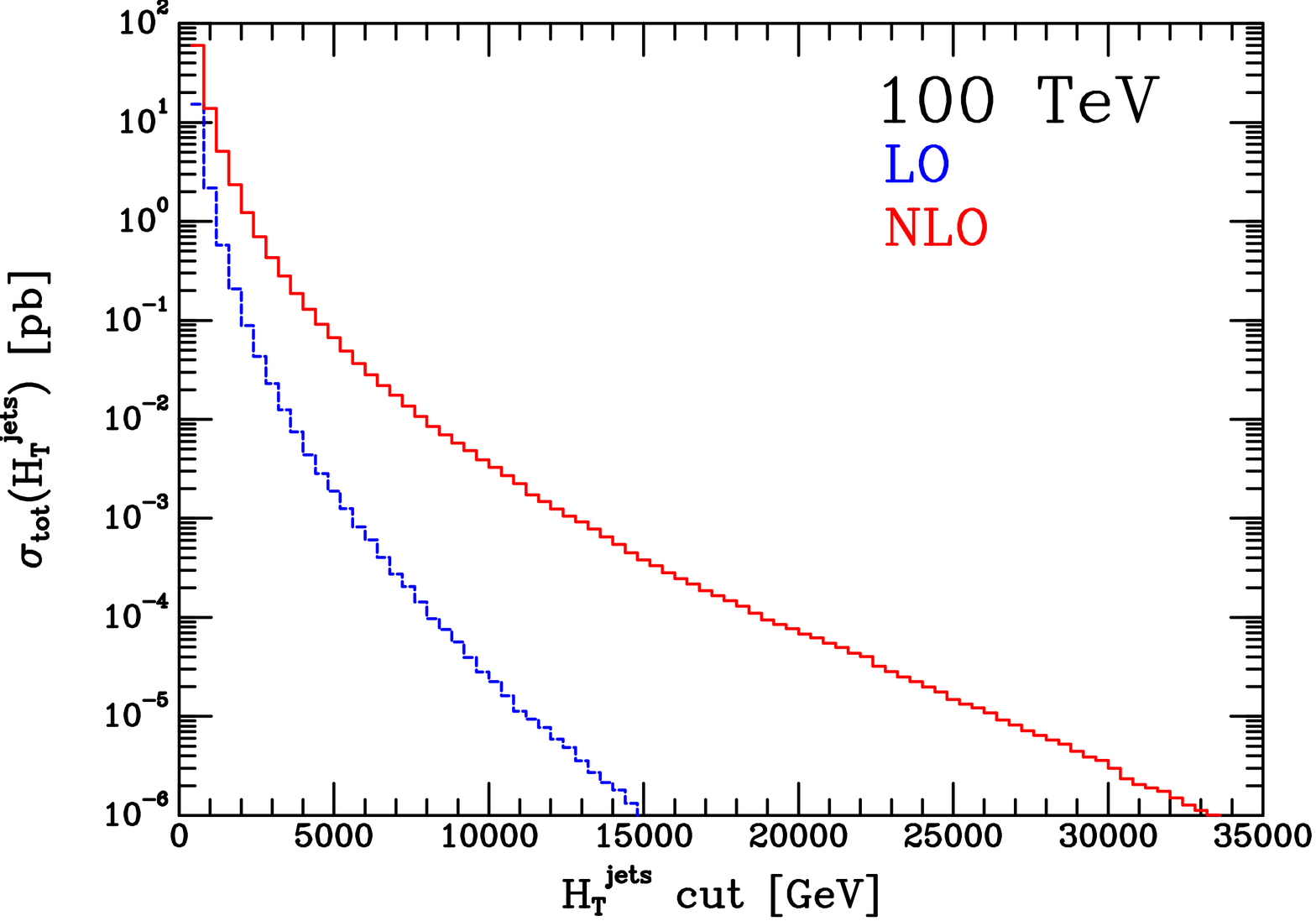} \hspace*{0.5cm}
\includegraphics[width=0.3\textwidth, angle=-90]{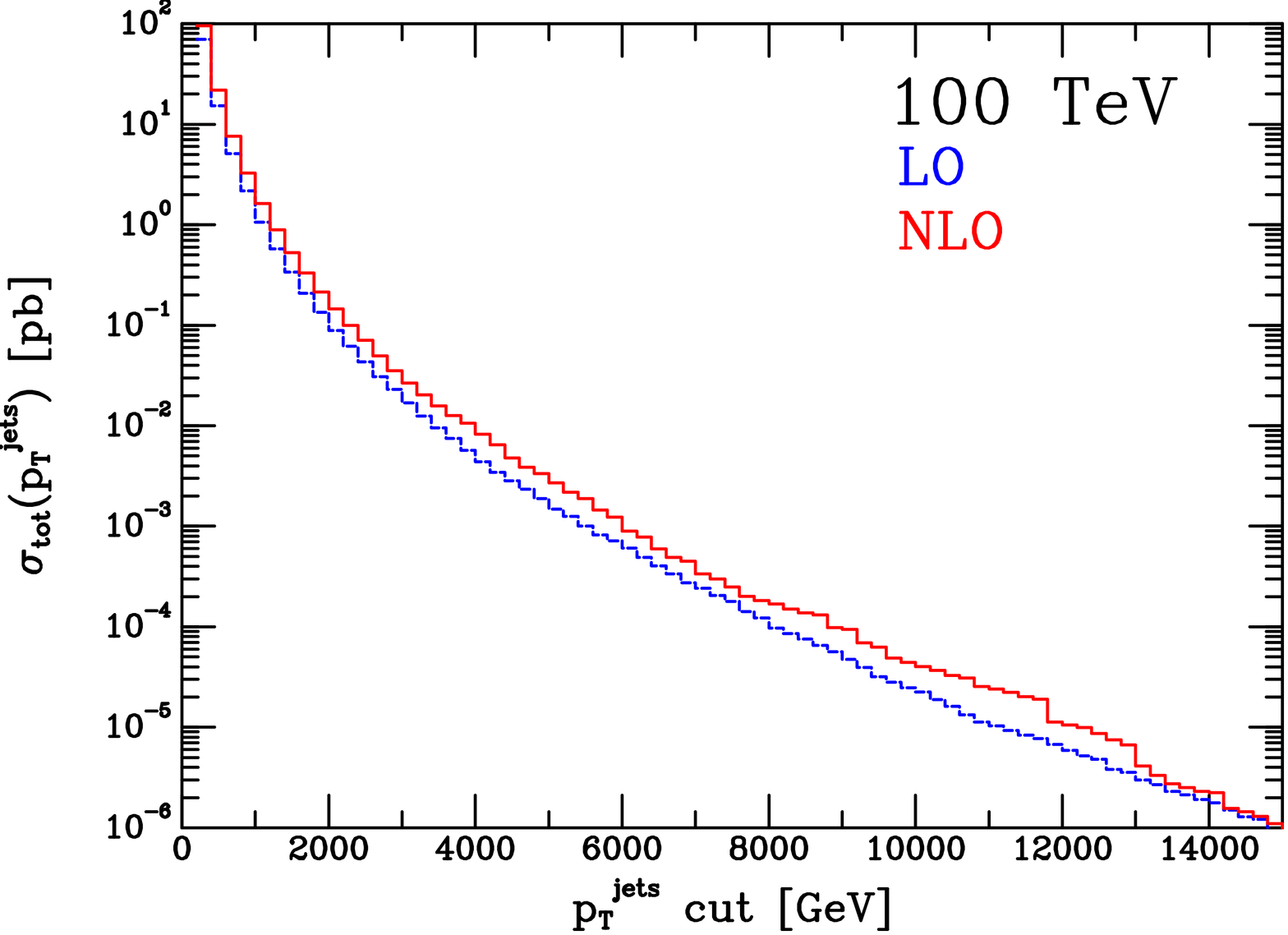}
\end{center}
\caption{Total integrated cross-sections at LO and NLO, with additional lower cuts on
$H_T^\text{jets}=\sum_\text{jets} p_\perp^\text{jet}$ (left) and $|p_T^{WW}|\,\equiv\,|p_T^\text{jets}|$. See main body of text for details.
\label{fig:xss}}
\end{figure}

Total cross-sections predicted at LO and NLO are shown in Fig.~\ref{fig:ptjet}, as a function
of $p_{\perp, \text{cut}}^{\text{jet}}$ and for values as large as $400$~GeV at the
$100$~TeV machine. All numbers cited in this section do not take into account the decays of the $W$ bosons, and branching ratios must be applied accordingly.
The theoretical uncertainty band is computed by using
a series of scale variations, cf. \cite{Campbell:2015hya,Campbell:2016fph}. The cross-sections at NLO are significantly larger than those at LO and, in
general, the uncertainty bands do not overlap.  At 100 TeV the cross-sections
are about an order of magnitude larger than at 14 TeV. For the case of a 100 TeV collider, we see that $p^{\text{jet}}_\perp$-cuts of $\mathcal{O}\lb 10\,\TeV \rb$ or higher still render measurable cross-sections, at least for a high-luminosity scenario, cf. Fig. \ref{fig:ptjet2}. Similar results are obtained for total cross-sections with an additional cut on either $H_T^\text{jets}=\sum_\text{jets} p_\perp^\text{jet}$ or $|p_T^{WW}|\,\equiv\,|p_T^\text{jets}|$, the transverse momentum of the {\sl complete} jet system, cf. Fig.~\ref{fig:xss}; in both cases, $p_{\perp,\text{cut}}^{\text{jet}}\,=\,25\,\GeV$. In case of the $p_\perp$ cut on the total jet system (or, equivalently, the $WW$ system), K-factors of $2-3$ prevail up to cut values $\lesssim\,16\,\TeV$.

\begin{figure}
\begin{center}
\includegraphics[width=0.33\textwidth,angle=-90]{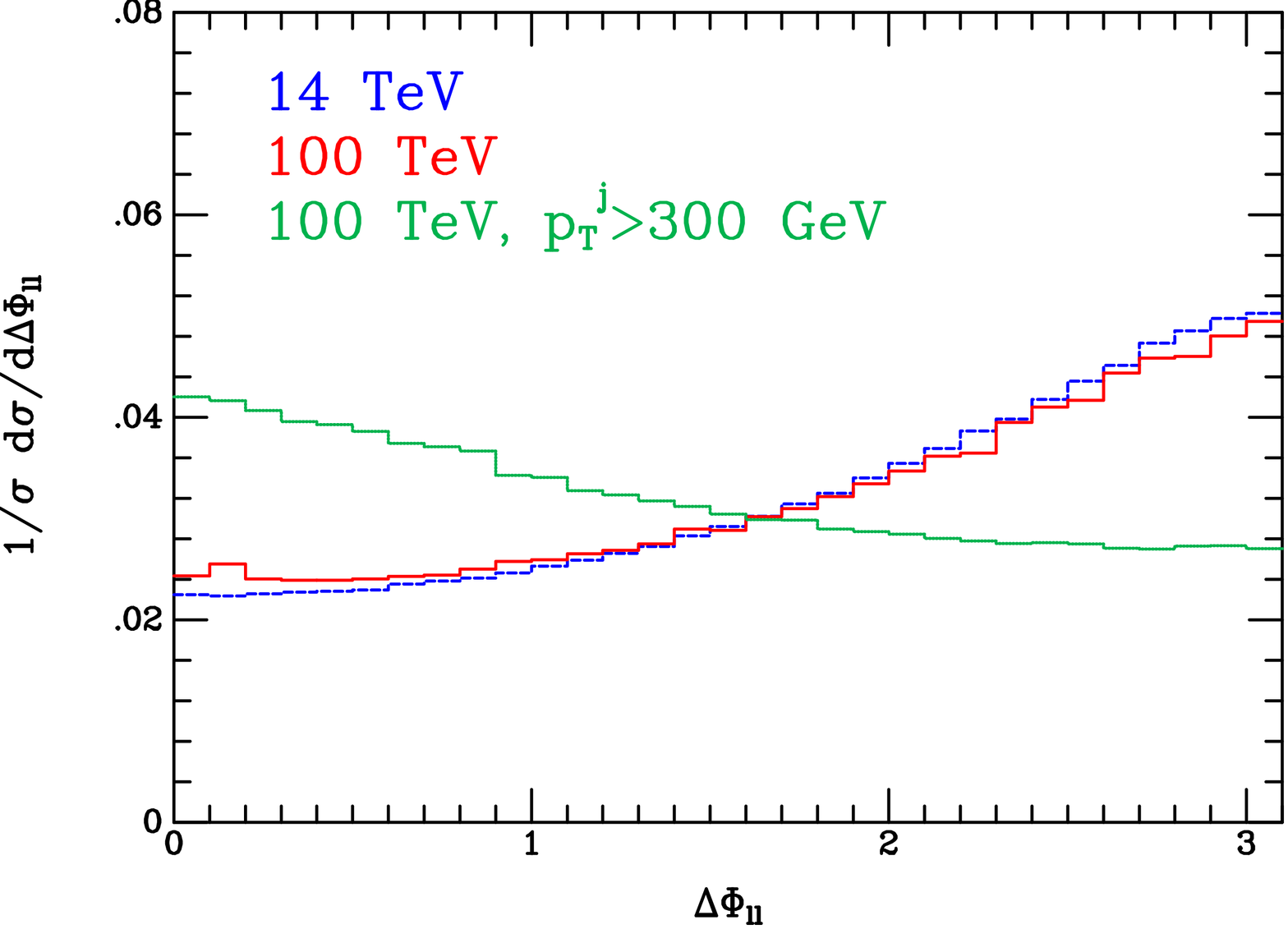} \hspace*{0.5cm}
\includegraphics[width=0.33\textwidth,angle=-90]{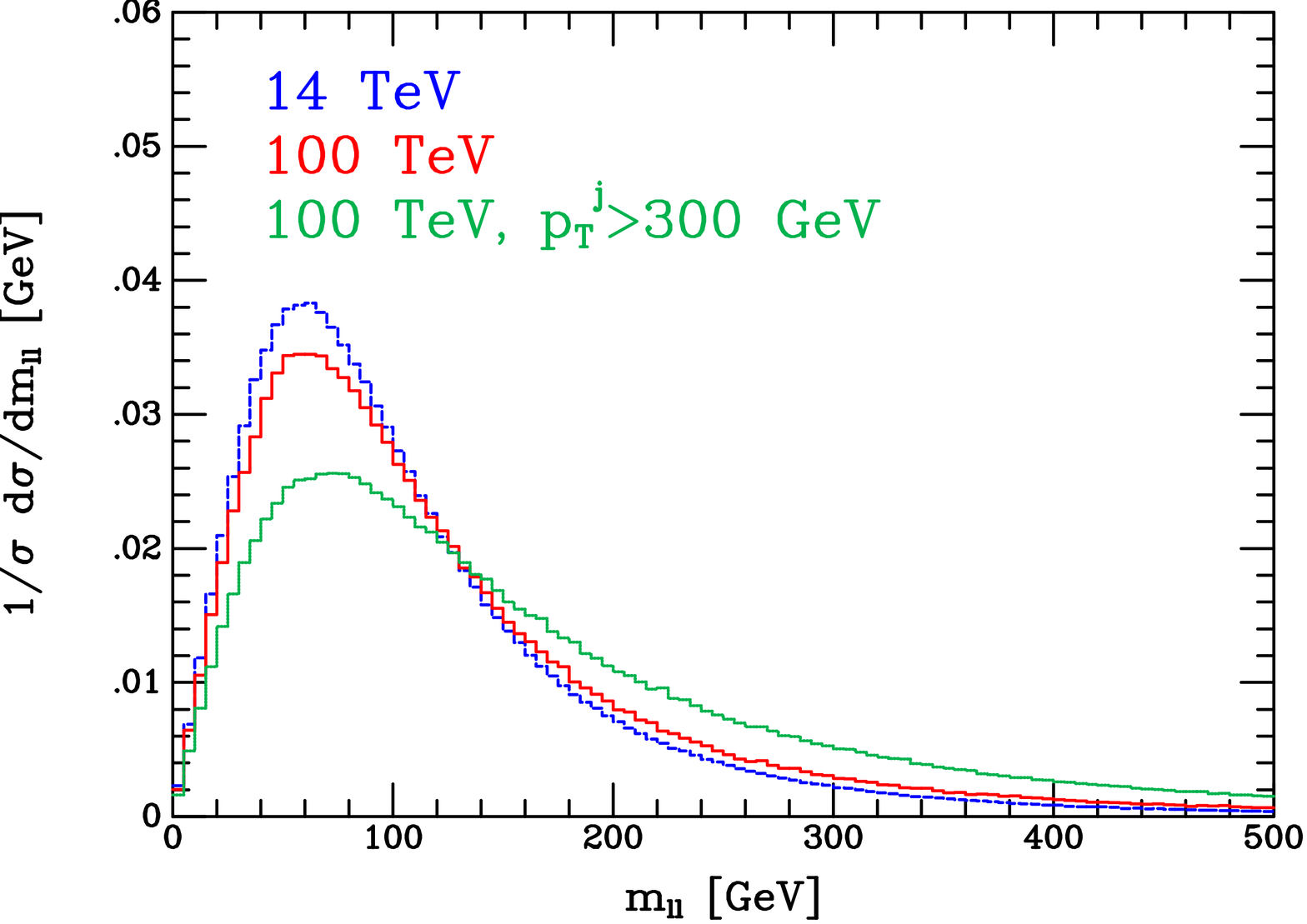}
\end{center}
\caption{NLO $\Delta \Phi_{\ell \ell}$ (left) and $m_{\ell \ell}$ (right)
distributions, normalized by the respective total cross-sections, for 14 \TeV (red),
100 \TeV (blue), and 100 \TeV with an increased $p_{\perp}^{\text{jet}}$ cut (green) 
\label{fig:comp3}}
\end{figure}

Regarding differential distributions, we briefly comment on an observable that is particularly interesting for Higgs searches at colliders, i.e. the azimuthal angle
between the electron and the positron, which can be used to isolate contributions to this
final state from Higgs boson decays. We here compare differential distributions  at the 14 TeV LHC as well as a 100 TeV collider, normalized by the respective total cross-section; in the latter case, we now additionally consider a scenario where the minimal $p_\perp$ cut on the jet has been set to 300 \GeV. Production cross-sections for these cases are given in Tab.~\ref{xsecs}. 
\renewcommand{\baselinestretch}{1.5}
\begin{table}
\begin{center}
\begin{tabular}{|r|r|c|c|}
\hline
$\sqrt s$~~~~ & $p_{\perp, \text{cut}}^{\text{jet}}$ & $\sigma_{LO}$~[pb] & $\sigma_{NLO}$~[pb] \\
\hline
$14$~TeV      &  25 GeV & $39.5_{-11.0\%}^{+11.7\%} $  & $48.6_{-4.0\%}^{+3.8\%}$ \\
$100$~TeV     &  25 GeV & $648_{-19.3\%}^{+22.3\%}   $  & $740_{-9.3\%}^{+4.5\%} $ \\
$100$~TeV     & 300 GeV & $30.3^{+11.22\%}_{-10.56\%}$&$53.7^{+8.0\%}_{-7.6\%}$ \\
\hline
\end{tabular}
\renewcommand{\baselinestretch}{1.0}
\caption{Cross-sections for the process $p p \to WW$+jet at proton-proton colliders
of various energies, together with estimates of the theoretical uncertainty
from scale variation as described in the text. 
Monte Carlo uncertainties are at most a
single unit in the last digit shown shown in the table.
\label{xsecs}}
\end{center}
\end{table}
As shown in Fig.~\ref{fig:comp3}, under the usual
jet cuts at $14$~TeV, this distribution is peaked towards $\Delta \Phi_{\ell \ell}=\pi$,
a feature which persists at $100$~TeV using the same jet cut.  Once the jet cut is raised
significantly, the recoil of the $W^+W^-$ system results in the two leptons instead being
preferentially produced closer together, i.e. in the region $\Delta \Phi_{\ell \ell} \to 0$; this region is usually favoured by events resulting from Higgs mediation.
Even if the jet threshold at a $100$~TeV collider were not as high as $300$~GeV, such a shift
in this distribution could be an important consideration in optimizing  the according analyses in this channel. 
On the other hand, the $m_{\ell \ell}$ distribution remains similar, albeit with a longer tail in the high-energy scenario.
\section{Summary}
In the current run of the LHC, precise knowledge of predictions for SM processes is more crucial than ever. We have considered the process $W^+\,W^-$ + jet at NLO QCD, making
use of an analytic calculation implemented into MCFM. We have considered total cross-sections as well as differential
distributions at proton-proton colliders with 14 TeV and 100
TeV center-of-mass energies for various cut scenarios. We found that in general at 100 TeV dimensionful variables
such as $m_{\ell\ell}$ exhibit longer tails in the distributions,
reflecting the increased center-of-mass energy of the system.
Furthermore, applying a higher $p_\perp$ cut significantly changes distributions
for the dilepton azimuthal angle $\Delta \Phi_{\ell\ell}$, frequently used for
background suppression for Higgs measurements.
\section*{Acknowledgements}
T.R. would like to thank the Fermilab Theory Group for their repeated hospitality
while this work was completed. DJM is supported by the UK Science and Technology Facilities Council (STFC) under grant ST/L000446/1.  This research is supported by the US DOE under contract DE-AC02-07CH11359.


\end{document}